\documentclass[12pt,a4paper,fleqn]{article}

\usepackage[textheight=23cm,textwidth=15cm]{geometry}
\usepackage{amsmath}
\usepackage{graphicx}
\usepackage[american]{babel}
\usepackage{here}
\usepackage[articletitle=true]{achemso}
\usepackage{url}

\begin{document}

\begin{center}

{\LARGE\bf
Ultra-Fast Semi-Empirical Quantum Chemistry for High-Throughput Computational Campaigns with Sparrow
}

\vspace{0.5cm}

{\large
Francesco Bosia$^{a,}$\footnote{ORCID: 0000-0001-6021-7672},
Peikun Zheng$^{b,}$\footnote{ORCID: 0000-0002-0248-936X},
Alain Vaucher$^{a,}$\footnote{ORCID: 0000-0001-7554-0288; present address: IBM Research Europe, 8803 R\"uschlikon, Switzerland},
Thomas Weymuth$^{a,}$\footnote{ORCID: 0000-0001-7102-7022},
Pavlo O.~Dral$^{b,}$\footnote{Corresponding author; e-mail: dral@xmu.edu.cn; ORCID: 0000-0002-2975-9876}, and
Markus Reiher$^{a,}$\footnote{Corresponding author; e-mail: markus.reiher@phys.chem.ethz.ch; ORCID: 0000-0002-9508-1565}
}\\[4ex]

$^{a}$ Laboratory of Physical Chemistry, ETH Zurich, Vladimir-Prelog-Weg 2, 8093 Zurich, Switzerland
\\
$^{b}$ State Key Laboratory of Physical Chemistry of Solid Surfaces, Fujian Provincial Key Laboratory of Theoretical and Computational Chemistry, Department of
Chemistry, and College of Chemistry and Chemical Engineering, Xiamen University, Xiamen 361005, China.

January 12, 2023

\vspace{0.5cm}

\textbf{Abstract}
\end{center}
\vspace{-0.7cm}
Semi-empirical quantum chemical approaches are known to compromise accuracy for feasibility of calculations on huge molecules. 
However, the need for ultrafast calculations
in interactive quantum mechanical studies, high-throughput
virtual screening, and for data-driven machine learning 
has shifted the emphasis towards calculation runtimes recently.
This comes with new constraints for the software implementation
as many fast calculations would suffer from a large overhead of
manual setup and other procedures that are comparatively fast
when studying a single molecular structure, but which become prohibitively slow
for high-throughput demands.
In this work, we discuss the effect of various well-established
semi-empirical approximations on calculation speed and relate this
to data transfer rates from the raw-data source computer to the 
results visualization front end. For the former, we consider desktop
computers, local high performance computing, as well as remote cloud services 
in order to elucidate the effect on interactive calculations, for web and cloud interfaces in local applications, and in world-wide interactive virtual sessions.
The models discussed in this work have been implemented into our
open-source software \textsc{SCINE Sparrow}.

\section{Introduction}
\label{sec:intro}
Semi-empirical methods have originated as feasible quantum chemical
approaches at times when computer hardware was struggling with advanced \textit{ab initio} methods (and even with Hartree--Fock calculations).
The advancement of modern computer hardware in the past three decades
has not rendered semi-empirical approaches superfluous, but instead pushed the
limits of molecular size that can be considered in a 
single electronic structure calculation (not requiring any embedding techniques, which would introduce further approximations of different flavor into the first-principles calculations). In recent years, several research groups have successfully carried out semi-empirical calculations for systems exceeding thousands of atoms. For example, Hennemann and Clark reported a calculation on no fewer than 100'000 atoms with their EMPIRE code\cite{Empire2014}. Dral presented results for carbon nanotubes and carbon peapods with about 1000 atoms\cite{Dral2014}. Thiel and coworkers developed a computer program accelerated by graphical processing units (GPUs), able to carry out calculations on systems with more than 1000 heavy atoms (\textit{e.g.}, a water cluster consisting of 1800 H\textsubscript{2}O molecules)\cite{MNDOGPU2012}. A similar approach was taken by Maia \textit{et al.}, who extended the MOPAC program to take advantage of GPUs. With the resulting software, these authors could also successfully carry out calculations for systems with more than 1000 atoms\cite{MOPACGPU2012}. Recently, Tretiak and coworkers described PYSEQM, a computer program for Born--Oppenheimer molecular dynamics based on semi-empirical methods; they presented results for a phenylene--ethynylene dendrimer consisting of 884 atoms\cite{PYSEQM}. Stewart demonstrated a calculation on a protein with 14'566 atoms\cite{Stewart2008}. Also the DFTB+ software package, implementing density functional tight binding methods, can be applied to systems with 1000 atoms and more\cite{Hourahine2020}.

Due to their short runtimes, semi-empirical methods allow one to conduct high-throughput virtual screening campaigns\cite{Shoichet2004, Pyzer-Knapp2015}
on very large molecule sets in comparatively little time. For example, Halls and Tasaki screened more than 7000 compounds for the suitability as lithium ion battery additives with the PM3 model\cite{Halls2010}. Jensen and coworkers presented a workflow based on the PM6 model to predict pK$_a$ values which is fast enough to be used in screening studies with as few as one hundred processors\cite{Kromann2016}. Zwijnenburg and coworkers proposed a high-throughput screening approach to assess optoelectronic properties of conjugated polymers which is based on the xTB model\cite{Wilbraham2018, Heath-Apostolopoulos2019}. Wen \textit{et al.}~assessed the detonation velocity of about 100'000 high-energy materials with the PM7 model\cite{Wen2022}.

In many cases, semi-empirical methods are employed in high-throughput screening studies as part of more complex workflows. For example, structures can be preoptimized with a fast semi-empirical calculation and then subsequently refined with a more accurate albeit computationally more costly method, or compound structures can be entirely determined with semi-empirical methods while some desired property is evaluated with a more accurate electronic structure model (see, \textit{e.g.}, Refs.~\citenum{Hachmann2011, Kanal2013, Korth2014, Martin2014, Cheng2015, Imamura2017, Schwobel2017, Irfan2022, QM9}).

The rise of data-driven machine learning approaches in computational chemistry also gives a substantial impetus for the application of semi-empirical methods, both for data generation and for developing new hybrid machine learning/semi-empirical methods. Data-driven approaches are hungry for data and fast semi-empirical methods allow for generating vast amounts of data within a reasonable amount of time (\textit{e.g.}, PM6 was used to calculate optimized molecular geometries and electronic properties of 221 million molecules\cite{PubChemQCPM6} and, earlier, PM7 was used to preoptimize structures of 134 thousand molecules in the QM9 dataset\cite{QM9}). Pure machine learning models are faster than semi-empirical methods but suffer from poorer transferability. The combination of machine learning with physics-based semi-empirical methods creates more transferable hybrid methods that can target higher accuracy approaches (mostly, density functional theory\cite{delta,MLDFTBYaron18,OrbNetDenali21,Frauenheim22,Abel22,Atz22,PNAS22}, but also a few methods\cite{MLOM2,AIQM1} such as AIQM1\cite{AIQM1} targeting coupled cluster or G4MP2 accuracy; see the recent overview in Ref.~\citenum{MLSQMreview23}). 
For instance, AIQM1 turned out to be computationally faster and at the same time more accurate than many standard density functional theory methods for various applications and even helped to revise experimental data\cite{AIQM1,AIQM1HoFs}.
 
An important area of application for ultrafast semi-empirical methods is interactive quantum mechanics\cite{Haag2013, Haag2014, Haag2014a, Weymuth2021}. Behind this approach is haptic quantum chemistry\cite{Marti2009} which exploits the tactile human sense so that a person can manipulate a molecule of interest in real time (\textit{e.g.}, try to abstract a proton) while the structure is continuously relaxed. For a realistic experience, it is important that this relaxation is carried out in a physically meaningful way, \textit{i.e.}, according to the laws of quantum mechanics. Furthermore, it needs to be fast enough so as not to break the interactivity. For a fluent visual feedback, about 30\,--\,60 calculations need to finish sequentially per second, while direct haptic (\textit{i.e.}, tactile) feedback requires, in principle, up to 1000 single-point calculations per second, which can, however, be reduced to a smaller number through activation of a mediator potential\cite{Vaucher2016}. Semi-empirical methods are generally fast compared to full-fledge electronic structure models, which makes them therefore ideally suited for interactive applications.

Last but not least, we note that very fast quantum chemical calculations are also a key ingredient to \textit{ab initio} molecular dynamics\cite{Meuwly2019}, where long trajectories with
femtosecond timestep resolution demand fast electronic structure programs to make thousands of single-point calculations viable.

Approaches to reduce the computational complexity of a given quantum chemical method, in particular linear-scaling approaches, obviously play an important role in the acceleration of quantum chemical calculations\cite{Ochsenfeld2007, Zalesny2011, Bowler2012, Mohr2015}. However, it is important to realize that such approaches are primarily targeting atomistic systems of increasing size. For small and medium-sized molecules, the overall calculation time might not be significantly reduced with respect to a canonical approach. This is an important observation, because especially for interactive applications the absolute calculation time is decisive rather than the relative scaling of a method. It is for this reason that our focus here will be on the actual, overall computation time of the methods discussed.
 
In this work, we report the implementation of a particular class of semi-empirical methods, the so-called orthogonalization-corrected methods (OMx)\cite{OM1_phd,OM1,OM2,OM2_phd,OM3_phd,OMx,ODMx}, as well as AIQM1, an artificial intelligence-enhanced method based on OMx, into our software package \textsc{SCINE Sparrow}\cite{sparrow, SQM_review_2018}. In addition, we investigate runtimes of various semi-empirical methods implemented in \textsc{Sparrow} and relate them to transmission times 
of various computer networks. In particular for interactive applications, these transmission times 
are an important factor to consider. This allows us to investigate the suitability of different combinations of semi-empirical methods and computing environments for high-throughput screening as well as interactive settings for various molecule sizes.
 
This work is organized as follows: In section~\ref{sec:theory}, we review the central approximations for the various classes of semi-empirical methods available in \textsc{Sparrow}. Particular emphasis is placed on the OMx methods. Then, we detail the computational methodology adopted in this work and present and analyze runtimes measured and relate them to the constraints set by interactive quantum mechanics and by high-throughput computational campaigns in section~\ref{sec:results}. Finally, we draw conclusions in section~\ref{sec:conclusion}.
 
\section{Theory}
\label{sec:theory}
In this section, we give a brief overview of the various semi-empirical methods implemented in \textsc{Sparrow} which are summarized in Table~\ref{tbl:summary}. For a more general introduction on semi-empirical methods we refer the reader to the recent review in Ref.~\citenum{SQM_review_2023}.

\begin{table}
\caption{Overview of the methods implemented in \textsc{Sparrow}.}
\centering
\begin{tabular}{llll} 
\hline
\hline
        Method    & Parametrized elements & Analytical gradients & Reference       \\ 
\hline
\hline
\multicolumn{4}{c}{Density functional tight binding} \\ \hline
DFTB0 & H, B\,--\,F, Si\,--\,Cl, Br, I, & yes & \citenum{Porezag1995, Seifert1996} \\
    &  Na, Mg, K\,--\,Ti, Fe\,--\,Ni, Zn\textsuperscript{a} & & \\
DFTB2 & H, B\,--\,F, Si\,--\,Cl, Br, I, & yes & \citenum{Elstner1998} \\
    &  Na, Mg, K\,--\,Ti, Fe\,--\,Ni, Zn\textsuperscript{a} & & \\
DFTB3 & H, B\,--\,F, Si\,--\,Cl, Br, I, & yes & \citenum{Gaus2011} \\
    &  Na, Mg, K\,--\,Ti, Fe\,--\,Ni, Zn\textsuperscript{a} & & \\
\hline
\multicolumn{4}{c}{Neglect of Diatomic Differential Overlap (MNDO-type)} \\ \hline
MNDO(/d) & H-Ca, Zn-Sr, In-Ba, Hg-Bi & yes & \citenum{MNDO,MNDOd} \\
AM1 & H-Be, C-Ca & yes & \citenum{AM1} \\
RM1 & H, C-F, P-Cl, Br, I & yes & \citenum{RM1} \\
PM3 & H-Ca, Zn-Sr, Cd-Ba, Hg-Bi & yes & \citenum{PM3_1,PM3_2} \\
PM6 & H-Bi except Ce-Lu\textsuperscript{b} & yes & \citenum{PM6}\\
\hline
\multicolumn{4}{c}{Neglect of Diatomic Differential Overlap} \\
\multicolumn{4}{c}{(orthogonalization-corrected and machine learning-enhanced)} \\ \hline
OM2 & H, C, N, O, F & no & \citenum{OM2,OM2_phd,OMx}\\
OM3 & H, C, N, O, F & no & \citenum{OM3_phd,OMx}\\
AIQM1 & H, C, N, O & no & \citenum{AIQM1}  \\
\hline
\hline
\multicolumn{4}{l}{\textsuperscript{a} Depending on the actual parameter set employed, parameters for some elements} \\
\multicolumn{4}{l}{will usually not be available in a given calculation (see Section~\ref{sec:dftb}).} \\
\multicolumn{4}{l}{\textsuperscript{b} For some pairs of elements some parameters are not available.} \\
\end{tabular}
\label{tbl:summary}
\end{table}

\subsection{Semi-Empirical Models and their Intricate Parameter Dependence}

\subsubsection{Neglect of Diatomic Differential Overlap}

The basic Hartree--Fock approach toward the solution of the nonrelativistic Schr\"odinger equation formally scales with the fourth power of the number of single-particle basis functions. This scaling was a severe bottleneck for the limited computational resources available in the 1960s and 1970s, which is why semi-empirical methods were developed by introducing rather drastic approximations.

An important class of semi-empirical models is based on the neglect of diatomic differential overlap (NDDO) that makes three key approximations. First, the number of explicitly considered electrons is reduced to a minimum. For example, for carbon, only four electrons are considered explicitly; the two $1s$-type core electrons are treated implicitly. In fact, the number of valence electrons considered explicitly is never larger than 12, for any element.

Second, the number of basis functions is reduced to the absolute minimum by relying on minimal basis sets. For example, for carbon, only an $s$-type and three $p$-type functions are used.

Third, the NDDO approximation itself is adopted to approximate the electron-repulsion integrals (ERIs),
\begin{equation}
\langle \phi_{\mu}\phi_{\nu}|\phi_{\lambda}\phi_{\sigma}\rangle \approx \delta_{\rm IJ}\delta_{\rm KL}\langle \chi_{\mu}^{\rm I}\chi_{\nu}^{\rm J}|\chi_{\lambda}^{\rm K}\chi_{\sigma}^{\rm L} \rangle.
\end{equation}
Here, $\phi_{\mu}$ is the $\mu$th orthogonalized basis function and $\chi_{\mu}^I$ is the $\mu$th atomic basis function centered on atom I. Any ERI for which basis functions $\mu$ and $\nu$ as well as $\lambda$ and $\sigma$ are not located on the same atom are neglected, hence lowering the scaling from the fourth to the second power in the number of basis functions. 

During the past decades a range of models has been developed based on these three core approximations. Notable examples are MNDO\cite{MNDO}, MNDO/d\cite{MNDOd}, AM1\cite{AM1}, RM1\cite{RM1}, PM3\cite{PM3_1, PM3_2}, PM6\cite{PM6}, and PM7\cite{PM7} as well as the family of orthogonalization-corrected methods described below.
For a more detailed introduction into the NDDO approximation and the multitude of models applying it, we refer the reader to the review in Ref.~\citenum{SQM_review_2018}.
\textsc{Sparrow} implements MNDO, AM1, RM1, PM3, and PM6\cite{Vaucher2018}. Note that extensions to PM6, such as PM6-DH\cite{Rezac2009}, PM6-DH2\cite{Korth2010}, PM6-DH+\cite{Korth2010a}, and PM6-D3H4\cite{Rezac2012}, which aim at improving the description of noncovalent interactions, are currently not available in \textsc{Sparrow}. For MNDO, parameters are supplied for elements H\,--\,Ca, Zn\,--\,Sr, In\,--\,Ba, and Hg\,--\,Bi. For AM1, the parameters cover elements H\,--\,Be, and C\,--\,Ca, while RM1 has parameters for H, C\,--\,F, P\,--\,Cl, Br, and I. PM3 features parameters for elements H\,--\,Ca, Zn\,--\,Sr, Cd\,--\,Ba, and Hg\,--\,Bi. PM6 covers all elements up to Bi except the lanthanides Ce\,--\,Lu. Note, however, that not every possible pairwise combination of these elements is parametrized. If parameters are missing for a certain element or element pair, \textsc{SCINE Sparrow} produces an error message, pointing out which parameters are exactly missing. Arbitrary, user-defined parameters can also be supplied.

\subsubsection{Density Functional Tight Binding}
\label{sec:dftb}

A different approach is taken by the so-called density functional tight binding (DFTB) family of models\cite{Elstner2014}. It is based on a Taylor series expansion of the electronic energy around a reference energy based on a superposition of atomic electron densities. With this ansatz, the electronic energy $E^{\rm DFTB}$ is obtained as
\begin{equation}
\label{eq:dftb}
E^{\rm DFTB} = E_{\rm rep} + \sum_{i}^{\rm occ}\epsilon_i + \frac{1}{2}\sum_{\rm A,B}^{N}\Delta q_{\rm A}\Delta q_{\rm B} \gamma_{\rm AB} + \frac{1}{3}\sum_{\rm A,B}^{N}\Delta q_{\rm A}^2\Delta q_{\rm B} \Gamma_{\rm AB} + \ldots
\end{equation}
Here, $E_{\rm rep}$ models the repulsion energy, whereas the second term represents the binding energy. The latter is based on a linear combination of atomic orbitals and determined from an eigenvalue problem similar to the one appearing in the standard Hartree--Fock model. In its simplest form (sometimes called DFTB0, non-self-consistent DFTB or non-SCC DFTB), only these first two terms are considered\cite{Porezag1995, Seifert1996}. Both terms are heavily parametrized, and the eigenvalue problem required for the second term does not have to be solved in an iterative fashion (hence the term ``non-self-consistent DFTB'').

The third term in Eq.~\eqref{eq:dftb} takes into account charge transfer, allowing for a better description of polarized systems. The charge fluctuations on atom A, $\Delta q_{\rm A}$, are determined in a Mulliken population analysis and $\gamma_{\rm AB}$ is a parametrized function of the internuclear distance between atoms A and B.
When this third term is also considered, the resulting method is called DFTB2 or self-consistent-charge DFTB (SCC-DFTB)\cite{Elstner1998}.

Finally, the DFTB3 method\cite{Gaus2011} takes into account density fluctuations up to third order by incorporating also the fourth term on the right-hand side of Eq.~\eqref{eq:dftb}. $\Gamma_{\rm AB}$ is also a parametrized function depending on the internuclear distance. Note that the function $\gamma_{\rm AB}$ is not exactly the same in DFTB3 compared to DFTB2, but contains an additional scaling factor.

Besides these three original DFTB approaches, another class of semi-empirical models based on the tight-binding idea was developed by Grimme and coworkers, namely GFN0-xTB\cite{Pracht2019}, GFN1-xTB\cite{Grimme2017}, and GFN2-xTB\cite{Bannwarth2019}. These models were developed with a focus on general applicability for almost all elements of the periodic table, which has been a key feature for their fast and widespread acceptance.

\textsc{Sparrow} implements DFTB0, DFTB2, and DFTB3\cite{Vaucher2018}. For these methods, the parameter sets 3ob-1-1 (covering elements C, H, N, and O), 3ob-2-1 (C, H, N, O, P, and S), 3ob-3-1 (covering C, H, N, O, P, S, F, Cl, Br, I, Na, K, Ca, Mg, and Zn), mio-1-1 (C, H, N, O, and S), mio-1-1 (containing C, H, N, O, P, and S), borg-0-1 (H, B), pbc-0-3 (C, H, N, O, Si, F, and Fe), trans3d-0-1 (C, H, N, O, Sc, Ti, Fe, Co, and Ni), and znorg-0-1 (covering C, H, N, O, S, and Zn) are supplied. Other, user-defined parameter sets can also be supplied.

\subsubsection{Orthogonalization-Corrected Methods}
For the present work, we have implemented in \textsc{Sparrow} orthogonalization-corrected methods (OMx) models: OM2\cite{OM2_phd,OM2}, OM3\cite{OM3_phd}, ODM2*\cite{ODMx}, ODM3*\cite{ODMx}, and AIQM1\cite{AIQM1}. Other OMx variants not implemented in \textsc{Sparrow} are OM1\cite{OM1_phd,OM1}, ODM2\cite{ODMx}, and ODM3\cite{ODMx}. Different from most traditional NDDO-based semi-empirical molecular orbital methods, such as MNDO\cite{MNDO}, AM1\cite{AM1}, PM3\cite{PM3_1,PM3_2}, PM6\cite{PM6}, and PM7\cite{PM7}, etc., OMx models include orthogonalization corrections, $V^{\text{ORT}}$, in the core Hamiltonian, $H^{\text{core}}$, which is part of the Fock matrix: 

\begin{equation} \label{OMx_1}
H_{\mu \nu}^{\text{core\,\,}}=U_{\mu \mu}\delta _{\mu \nu}+\sum_{\text{B}}{\left[ V_{\mu \nu ,\text{B}}^{\text{s}}+V_{\mu \nu ,\text{B}}^{\text{ORT}}+V_{\mu \nu ,\text{B}}^{\text{PI}}+V_{\mu \nu ,\text{B}}^{\text{ECP}} \right]},
\end{equation}
\begin{equation} \label{OMx_2}
H_{\mu \lambda}^{\text{core\,\,}}=\beta _{\mu \lambda}+\sum_{\text{C}}{V_{\mu \lambda ,\text{C}}^{\text{ORT}}}.
\end{equation}

In addition, the attractive penetration integrals, $V^{\text{PI}}$, and the repulsive effective core potentials, $V^{\text{ECP}}$ are also incorporated into the one-center core Hamiltonian matrix.
Here, we follow a common notation to distinguish different centers of basis functions: subscripts $\mu$ and $\upsilon$ denote basis functions centered on the same atom A, while subscript $\lambda$ denotes basis functions at another atom B, which is different from A, and C represents an atom which is neither A, nor B.

The first term $U_{\mu\mu}$ in Eq.~\eqref{OMx_1} is the one-center one-electron energy of atom A and is obtained by parameterization. $V_{\mu\upsilon, B}^s$ is the semi-empirical core--electron attraction and calculated as in MNDO-type models:

\begin{equation} \label{OMx_3}
V_{\mu \nu ,\text{B}}^{\text{s}}=-Z_{\text{B}}\left( \mu ^{\text{A}}\nu ^{\text{A}},s^{\text{B}}s^{\text{B}} \right) ^{\text{s}},
\end{equation}

where $Z_{\text{B}}$ is the core charge of atom B, and the semi-empirical two-center electron ERIs are obtained by multiplying the analytic ERIs 
$\left( \mu ^{\text{A}}\nu ^{\text{A}},s^{\text{B}}s^{\text{B}} \right) ^{\text{s}}$
with a Klopman–Ohno scaling factor $f_{\text{KO}}$:

\begin{equation} \label{OMx_4}
\left( \mu ^{\text{A}}\nu ^{\text{A}},\lambda ^{\text{B}}\sigma ^{\text{B}} \right) ^{\text{s}}=f_{\text{KO}}\left( \mu ^{\text{A}}\nu ^{\text{A}},\lambda ^{\text{B}}\sigma ^{\text{B}} \right) ^{\text{a}}.
\end{equation}

The introduction of $f_{\text{KO}}$ ensures that the semi-empirical two-center ERIs smoothly transition to the corresponding values in the one-center limit. It is given by

\begin{equation} \label{OMx_5}
f_{\text{KO}}=\left( s^{\text{A}}s^{\text{A}},s^{\text{B}}s^{\text{B}} \right) ^{\text{MNDO}}/\left( s^{\text{A}}s^{\text{A}},s^{\text{B}}s^{\text{B}} \right) ^{\text{a}},
\end{equation}

where the integrals in the numerator are the MNDO-type ERIs and calculated from semi-empirical multipole-multipole interactions.
The core-core repulsion energy is also scaled with the $f_{\text{KO}}$ factor,

\begin{equation} \label{OMx_6}
E_{\text{AB}}^{\text{core\,\,}}=f_{\text{KO}}Z_{\text{A}}Z_{\text{B}}/R_{\text{AB}},
\end{equation}

where $R_{\text{AB}}$ is the distance between atoms A and B.

Another key feature of OMx methods is that the minimal valence contracted Gaussian type orbitals (GTOs) instead of Slater type orbitals (STOs) are employed in integral calculations, and a scaling factor $\zeta$ is used to adjust the exponents of the primitive Gaussian functions. That is, an STO-3G contraction is used for hydrogen and ECP-3G for all other elements. Then, the (analytic) ERIs -- and also the overlap integrals below -- can be evaluated analogously to the full ``\textit{ab initio}'' ones with the \textsc{Libint} library\cite{libint} in \textsc{Sparrow}. 

In the following, we provide a detailed description of $V^{\text{ORT}}$, $V^{\text{PI}}$, and $V^{\text{ECP}}$ because these three terms are the key ingredients in OMx models (beyond MNDO-type models) and achieve systematic improvements. Moreover, their description is scattered over multiple resources in the original literature, which may hamper the implementation of OMx models.

\textbf{Orthogonalization corrections $V^{\text{ORT}}$.}
The inclusion of the orthogonalization correction $V^{\text{ORT}}$ retrieves the effect of a transformation from a nonorthogonal (``atomic-orbital'') basis \{$\chi$\} to an orthogonal (``molecular-orbital'') basis \{$\phi$\} by expanding the Löwdin orthogonalization\cite{Lowdin1950} to second order in terms of overlap integrals:
\begin{equation} \label{LD_1}
\begin{aligned}
{}^\phi \boldsymbol{H}=\left({}^{\chi} \boldsymbol{S}\right)^{-\frac{1}{2}} {}^{\chi} \boldsymbol{H}\left({ }^{\chi} \boldsymbol{S}\right)^{-\frac{1}{2}}
=\left(\mathbf{1}+{}^{\chi} \boldsymbol{S}^{\prime}\right)^{-\frac{1}{2}}{}^{\chi} \boldsymbol{H}\left(\mathbf{1}+{}^{\chi} \boldsymbol{S}^{\prime}\right)^{-\frac{1}{2}},
\end{aligned}
\end{equation}

\begin{equation} \label{LD_2}
\begin{aligned}
{}^\phi \boldsymbol{H} \approx &{}^\chi \boldsymbol{H}-\frac{1}{2}\left({}^\chi \boldsymbol{S}^{\prime} {}^\chi \boldsymbol{H}+{ }^\chi \boldsymbol{H}{ }^\chi \boldsymbol{S}^{\prime}\right)+\frac{3}{8}\left({ }^\chi \boldsymbol{S}^{\prime 2}{ }^\chi \boldsymbol{H}+{ }^\chi \boldsymbol{H}^\chi \boldsymbol{S}^{\prime 2}\right)\\
&+\frac{1}{4}{ }^\chi \boldsymbol{S}^{\prime}{ }^\chi \boldsymbol{H}^\chi \boldsymbol{S}^{\prime}+\mathcal{O}\left({ }^\chi \boldsymbol{S}^{\prime 3}\right),
\end{aligned}
\end{equation}
where ${}^\chi\boldsymbol{S}^{\prime}={}^\chi\boldsymbol{S}-\mathbf{1}$.

Based on Eq.~\eqref{LD_2}, OMx methods adopt different parametric formulae in the the matrix-element calculation of ${}^\phi \boldsymbol{H}$. For simplicity, we will neglect the superscripts $\chi$ and $\phi$ in the following.

The very first OMx method OM1 only introduces the one-center orthogonalization corrections $V_{\mu\nu,\mathrm{B}}^{\mathrm{ORT}}$ defined as
\begin{equation} \label{OMx_7}
\begin{aligned}
	V_{\mu \nu ,\text{B}}^{\text{ORT}}=&-\frac{1}{2}F_{1}^{\text{A}}\sum_{\rho \in \text{B}}{\left( S_{\mu \rho}\beta _{\rho \nu}+\beta _{\mu \rho}S_{\rho \nu} \right)}\\
	&+\frac{1}{8}F_{2}^{\text{A}}\sum_{\rho \in \text{B}}{S_{\mu \rho}}S_{\rho \nu}\left( H_{\mu \mu ,\text{B}}+H_{\nu \nu ,\text{B}}-2H_{\rho \rho ,\text{A}} \right),
\end{aligned}
\end{equation}
where $F_1$ and $F_2$ are element-dependent parameters controlling the magnitudes of the one-center orthogonalization corrections. $S_{\mu\rho}$ and $\beta_{\rho\nu}$ are the overlap and resonance integrals, respectively.

The resonance integrals represent the two-center one-electron integrals $\left \langle\phi_\mu^A|T+V_A+V_B|\phi_\mu^B\right \rangle$ containing kinetic energy and nuclear--electron attraction terms arising from orbitals $\phi_\mu^A$ and $\phi_\mu^B$ on atoms A and B. Due to the severe effect of symmetric orthogonalization\cite{NDDO_analysis}, they are not evaluated analytically, but by an empirical formula in attempt to recover the orthogonalization effects through parametrization. The empirical formula is different compared to MNDO-based methods and given by\cite{OM1_phd,OM1}
\begin{equation} \label{OMx_8}
\beta _{\mu \lambda}=\frac{1}{2}\left( \beta _{\mu}^{\text{A}}+\beta _{\lambda}^{\text{B}} \right) \sqrt{R_{\text{AB}}}\exp \left[ -\left( \alpha _{\mu}^{\text{A}}+\alpha _{\lambda}^{\text{B}} \right) R_{\text{AB}}^{2} \right],
\end{equation}
where $\alpha^\text{X}$ and $\beta^\text{X}$  are element- and orbital-type-dependent parameters. The resonance integrals are first calculated in a local diatomic coordinate system and then transformed into the global coordinate system.

For $s$, $p_x$, $p_y$, $p_z$ type basis functions, the rotation matrices for this transformation from the local to the global coordinate system have the following form:
\begin{equation} \label{OMx_9}
\boldsymbol{M}^\text{rot}=\left[ \begin{matrix}
	1&		0&		0&		0\\
	0&		\sin b\cos a&		\sin b\sin a&		\cos b\\
	0&		\cos b\cos a&		\cos b\sin a&		-\sin b\\
	0&		-\sin a&		\cos a&		0\\
\end{matrix} \right],
\end{equation}
where the angle $b$ is defined as the angle between vectors $\boldsymbol{z}$ and $\boldsymbol{z}^{\text{loc}}$, and the angle $a$ is the angle between vector $\boldsymbol{x}$ and the projection of $\boldsymbol{z}^{\text{loc}}$ into the $xy$ plane. The details of the  construction of vectors $\boldsymbol{x}^{\text{loc}}$, $\boldsymbol{y}^{\text{loc}}$, $\boldsymbol{z}^{\text{loc}}$ can be found in the Supporting Information of Ref.~\citenum{SQM_review_2018}.

Considering the integral elements of matrix $\boldsymbol{I}$ in the local coordinate systems of atoms A and B, their corresponding values in the global coordinate system can be calculated by:
\begin{equation} \label{ROT}
\boldsymbol{I}=\begin{cases}
	\boldsymbol{I}^{\text{loc}},&		n_{\text{A}}=n_{\text{B}}=1,\\
	\left( \boldsymbol{M}^{\text{rot}} \right) ^T\boldsymbol{I}^{\text{loc}},&		n_{\text{A}}=4,n_{\text{B}}=1,\\
	\boldsymbol{I}^{\text{loc}}\boldsymbol{M}^{\text{rot}},&		n_{\text{A}}=1,n_{\text{B}}=4,\\
	\left( \boldsymbol{M}^{\text{rot}} \right) ^T\boldsymbol{I}^{\text{loc}}\boldsymbol{M}^{\text{rot}},&		n_{\text{A}}=n_{\text{B}}=4,\\
\end{cases}
\end{equation}
where $n_\text{A}$ and $n_\text{B}$ are the number of basis functions on atoms A and B, respectively.

The matrix element $H_{\mu\mu,\mathrm{X}}$ is restricted to an atom pair A--X and is calculated by
\begin{equation} \label{OMx_10}
H_{\mu \mu ,\text{X}}=U_{\mu \mu}+V_{\mu \mu ,\text{X}}^{\text{s}}.
\end{equation}

It should be noted that the second sum in Eq.~\eqref{OMx_7} is implemented by calculating each contribution in a local coordinate system. Hence, a local to global coordinate transformation is required.

In contrast to OM1, OM2 additionally includes two-center orthogonalization corrections $V_{\mu\lambda,\mathrm{C}}^{\mathrm{ORT}}$ in the core Hamiltonian in Eq.~\eqref{OMx_2}: 
\begin{equation} \label{OMx_11}
\begin{aligned}
	V_{\mu \lambda ,\text{C}}^{\text{ORT}}=&-\frac{1}{2}G_{1}^{\text{AB}}\sum_{\rho \in \text{C}}{\left( S_{\mu \rho}\beta _{\rho \lambda}+\beta _{\mu \rho}S_{\rho \lambda} \right)}\\
	&+\frac{1}{8}G_{2}^{\text{AB}}\sum_{\rho \in \text{C}}{S_{\mu \rho}}S_{\rho \lambda}\left( H_{\mu \mu ,\text{C}}+H_{\lambda \lambda ,\text{C}}-H_{\rho \rho ,\text{A}}-H_{\rho \rho ,\text{B}} \right),\\
\end{aligned}
\end{equation}
where $G_1^{\mathrm{AB}}$ and $G_2^{\mathrm{AB}}$ are defined as the arithmetic means of element-dependent parameters $G$:
\begin{equation} \label{OMx_12}
G_{1}^{\text{AB}}=\frac{1}{2}\left( G_{1}^{\text{A}}+G_{1}^{\text{B}} \right),
\end{equation}
\begin{equation} \label{OMx_13}
G_{2}^{\text{AB}}=\frac{1}{2}\left( G_{2}^{\text{A}}+G_{2}^{\text{B}} \right).
\end{equation}

All terms in Equation~\eqref{OMx_11} are directly calculated in the global coordinate system. To ensure rotational invariance, $H_{\mu\mu,\mathrm{X}}$ needs to be averaged when $\mu$ is a $p$-type basis function
\begin{equation} \label{OMx_14}
H_{pp,\text{X}}=\frac{1}{3}\left( H_{xx,\text{X}}+H_{yy,\text{X}}+H_{zz,\text{X}} \right).
\end{equation}
Differing from OM2, OM3 neglects the second-order terms in one- and two-center orthogonalization corrections, \textit{i.e.}, both $F_2^\text{X}$ and $G_2^\text{X}$ are zeros in OM3.

\textbf{Penetration integrals $V^{\text{PI}}$.}
The penetration integrals $V_{\mu\nu,\mathrm{B}}^{\mathrm{PI}}$ are defined as the difference between the actual core--electron attraction integrals, which are obtained from the analytic core--electron attraction integrals $V_{\mu\nu,\mathrm{B}}^\mathrm{a}$ scaled by $f_{\mathrm{KO}}$, and their corresponding semi-empirical analogs:
\begin{equation} \label{OMx_15}
V_{\mu \nu ,\text{B}}^{\text{PI}}=f_{\text{KO}}V_{\mu \nu ,\text{B}}^{\text{a}}-V_{\mu \nu ,\text{B}}^{\text{s}}.
\end{equation}
In our \textsc{Sparrow} implementation, $V_{\mu\nu,\mathrm{B}}^\mathrm{a}$ are evaluated by the integral engine of \textsc{Libint}. Note that there is a slight mistake in the description of the OMx methods in Ref.~\citenum{SQM_review_2018}: For the OMx methods, the first two terms of the second sum of Eq.~(77) of Ref.~\citenum{SQM_review_2018} always cancel each other. Accordingly, Eq.~(80) of Ref.~\citenum{SQM_review_2018} is always fulfilled, not only for s-type orbitals.

\textbf{Effective core potentials $V^{\text{PI}}$.}
To account for core--valence interactions, ECP contributions are incorporated into OMx methods. In OM2 and OM3, the following empirical expression is used to calculate $V_{\mu\nu,\mathrm{B}}^{\mathrm{ECP}}$:

\begin{equation} \label{OMx_16}
V_{\mu \nu ,\text{B}}^{\text{ECP}}=-\sum_{\alpha \in \text{B}}{\left( S_{\mu \alpha}G_{\alpha \nu}+G_{\mu \alpha}S_{\alpha \nu}+S_{\mu \alpha}S_{\alpha \nu}F_{\alpha \alpha} \right)},
\end{equation}

where $F_{\alpha\alpha}$ is treated as an atomic parameter and $G_{\mu\alpha}$ is represented by an expression analogous to the resonance integrals:

\begin{equation} \label{OMx_17}
G_{\mu \alpha}=\frac{1}{2}\left( \beta _{\mu}^{\text{A}}+\beta _{\alpha}^{\text{B}} \right) \sqrt{R_{\text{AB}}}\exp \left[ -\left( \alpha _{\mu}^{\text{A}}+\alpha _{\alpha}^{\text{B}} \right) R_{\text{AB}}^{2} \right].
\end{equation}

Two further atomic parameters, $\alpha_\alpha$ and $\beta_\alpha$, have been introduced here. 

The ODM2 and ODM3 models employ the same electronic structure model as OM2 and OM3. Compared with OM2 and OM3, however, Grimme's D3 dispersion corrections\cite{Grimme2010} are included as an integral part in ODMx. In addition to these dispersion corrections, a much more robust parametrization procedure and a wide range of training sets (including noncovalent interactions and electronic excitation energies) were used in the parametrization of ODMx models. Moreover, the calculation of enthalpies of formation in ODMx is analogous to \textit{ab initio} methods by explicitly adding zero-point vibrational energy and thermal corrections while other NDDO-based methods derive enthalpies of formation directly from self-consistent field ( SCF) total electronic energies. In \textsc{Sparrow}, we have not implemented the two- and three-body D3 dispersion corrections yet and, hence, our ODMx implementation does not contain any dispersion corrections. 
We denote the resulting methods as ODM2* and ODM3*, which is consistent with notation introduced previously\cite{AIQM1}. The Hamiltonians of the ODMx* models are then the same as in the OMx methods but the ODMx* models have different parameters trained on a more diverse data set in addition to the aforementioned difference in the treatment of SCF total electronic energies.

Up to now, OMx models have only been parameterized for second-period elements H, C, N, O, and F. Moreover, it must be emphasized that there are some slight differences in physical constants used for MNDO-type (constants taken from CODATA2014\cite{Mohr2016})
and for OMx methods in \textsc{Sparrow} (Table~\ref{tbl:consnts}) 
to ensure that the results are consistent with the original implementation of OMx methods.

\begin{table}
\caption{The physical constants applied in MNDO-type and OMx models.}
\centering
\begin{tabular}{lll} 
\hline
\hline
            & MNDO-type methods                                  & OMx                                            \\ 
\hline
Bohr radius & $0.52917721067\cdot10^{-10}$ m        & $0.529167\cdot10^{-10}$  m              \\ 
Hartree/eV  & $3.674932248\cdot10^{-2}$            & 1/27.21   \\ 
 & & $\approx 3.6751194414\cdot10^{-2}$ \\
eV/Hartree  & $1/3.674932248\cdot10^{-2}$ & 27.21                                          \\
 & $\approx27.21138602$ & \\
\hline
\hline
\end{tabular}
\label{tbl:consnts}
\end{table}

\subsubsection{AIQM1}
The AIQM1 method is an artificial-intelligence-enhanced quantum mechanical method, which can be used out of the box for very fast quantum chemical calculations with an accuracy as that of the 'gold-standard' coupled cluster approach for ground-state energies of neutral, closed-shell species and with reasonable accuracy for other properties, charge, and spin states\cite{AIQM1}. The energy of AIQM1 is composed of three parts: the baseline energy is obtained with the ODM2* Hamiltonian, then there is an ANI-type neural network (NN) correction\cite{ANI-1, ANI-1ccx, ANI-2x}, and a D4 dispersion correction\cite{D4_1,D4_2}: 
\begin{equation} \label{EAIQM1}
E_{\text{AIQM1}} = E_{\text{ODM2*}} + E_{\text{NN}} + E_{\text{D4}}.
\end{equation}
Since NN corrections are added on top of a prediction with a semi-empirical model, AIQM1 can be considered as a universal $\Delta$-learning model~\cite{delta} which is improved by transfer learning: first, NN corrections were trained on 4.6~million data points measuring the difference between density functional theory and ODM2* energies and forces. Then, NNs were partially refitted to a smaller 0.5~ million data points set of coupled clsuter and ODM2* (+ dispersion correction) energy differences. The NN correction $E_{\text{NN}}$ is calculated by using the locality approximation introduced by Behler and Parrinello\cite{BPNN}, \textit{i.e.}, it is the sum of atomic energy contributions $E_i$:
\begin{equation} \label{BPNN}
E_{\text{NN}} = \sum_i^{N_{\text {atom }}} E_i.
\end{equation}
The NN part of AIQM1 consists of an ensemble of eight ANI-type NNs\cite{TorchANI}, and the architecture of each NN in AIQM1 is similar to those of ANI-1x\cite{ANI-1x} and ANI-1ccx\cite{ANI-1ccx} except for two modifications: the previous CELU (continuously differentiable exponential linear unit) activation function was changed to GELU (Gaussian error linear unit) to ensure that higher derivatives can be obtained with this NN; the angular cutoff was increased from 3.5 to 4\,$\text {\AA}$.

Importantly, an ensemble of eight NNs not only improves overall accuracy of an approach but enables convenient uncertainty quantification (UQ) of AIQM1 predictions by evaluating standard deviations in NN corrections (NN SDs)\cite{AIQM1HoFs}. Such mechanism of UQ is lacking for any other traditional semi-empirical method not based on machine learning. We found that if NN SDs do not exceed 0.41\,kcal/mol, AIQM1 enthalpies of formation are generally reaching chemical accuracy (errors lower than 1\,kcal/mol)\cite{AIQM1HoFs}.

The dispersion energy $E_{\rm disp}$ is obtained from a sum over all pairs AB of atoms A and B,
\begin{equation}
 E_{\rm D4} = -\sum_{\rm A, B>A}\left( s_6\frac{C_6^{\rm AB}}{R_{\rm AB}^6}f_{\rm damp}^{(6)}(R_{\rm AB}) + s_8\frac{C_8^{\rm AB}}{R_{\rm AB}^8}f_{\rm damp}^{(8)}(R_{\rm AB}) \right) + E^{\rm ABC}~.
\end{equation}
The scaling factors $s_6$ and $s_8$ are empirical parameters. The dispersion coefficients $C_n^{\rm AB}$ ($n = 6, 8$) are obtained from tabulated values\cite{D4_1}, 
the Mulliken charge and the coordination numbers of atoms A and B. The damping function $f_{\rm damp}^{(n)}(R_{\rm AB})$ is a parametrized function of the internuclear distance $R_{\rm AB}$\cite{Grimme2011}.
The D4 dispersion corrections used in AIQM1 include Axilrod–Teller–Muto three-body contributions,
\begin{equation}
  E^{\rm ABC} = \frac{C_9^{\rm ABC}\left( 3{\rm cos}\theta_{\rm A} {\rm cos}\theta_{\rm B} {\rm cos}\theta_{\rm C} + 1\right)}{(R_{\rm AB}R_{\rm BC}R_{\rm CA})^3},
\end{equation}
where the triple-dipole constant $C_9^{\rm ABC}$ is approximated as the geometric mean of the $C_6$ coefficients of atom pairs AB, AC, and BC, and $\theta_{\rm A}$, $\theta_{\rm B}$, and $\theta_{\rm C}$ are the angles of the triangle formed by $R_{\rm AB}$, $R_{\rm BC}$, and $R_{\rm CA}$.

In our implementation, the $E_{\text{NN}}$ and $E_{\text{D4}}$ parts are calculated by interfacing to the \textsc{TorchANI}\cite{TorchANI} and \textsc{dftd4}\cite{dftd4} programs, respectively. The current parametrization of AIQM1 only supports the ``standard'' elements C, H, N, and O.

\subsection{Constraints on Interactivity and High Throughput}

Interactive and high-throughput calculations have the common requirement that calculation times should be as short as possible.
There is a variety of ways to accelerate quantum chemical calculations (see, \textit{e.g.}, Ref.~\citenum{Haag2013}). On the one hand, one can speed up the calculation for a given system size by relying on highly optimized computer programs, by parallelizing time-consuming parts of the calculation, and by taking advantage of sophisticated algorithms (\textit{cf.}, the multitude of methods available to accelerate SCF convergence\cite{Saunders1973, Pulay1980, Pulay1982, Kudin2002, Host2008, Hu2010, Chen2011, Wang2011, Muhlbach2016, Atsumi2010}). 
On the other hand, one can try to reduce the system size (\textit{i.e.}, by reducing or simplifying the Fock matrix while keeping the molecule identical). This can be achieved, for example, with standard integral screening approaches\cite{Almlof1982, Cremer1986, Haser1989}, and by using small basis sets. For interactive quantum chemistry, ultra-fast semi-empirical methods, which make use of (almost) all these approaches, are currently state of the art.

For interactive calculations, two more aspects are decisive for a successful application, namely the latency and bandwidth available for the transmission 
of the results to operators. Latency is a general term describing the time interval between the cause and effect of some physical change in a system. It is ultimately due to the limited velocity with which any physical interaction can propagate. Here, we use it to describe the time it takes for the results of a finished calculation to be transmitted 
to a user. On a standard local ethernet network, typical latencies are below 1\,ms, but on the internet, latencies on the order of 10\,ms and more may occur if packets are routed via several devices (such as routers or switches). Bandwidth refers to the amount of data which can be transmitted to the user per unit time. On a local network, bandwidths of 1\,Gb/s can be easily achieved. For a large-scale network such as the internet, the practically available bandwidth is typically lower, often on the order of 100\,Mb/s.

It is straightforward to see that the bandwidth is not a problem for interactive calculations. For a basic interactive exploration of a chemical system, the energy and the nuclear gradient (\textit{i.e.}, the forces acting on all atoms) must be calculated and transmitted. 
While the energy is a single floating-point number, the nuclear gradient consists of $3N$ such numbers, with $N$ being the number of atoms present. Even for a comparatively large molecule of 100 atoms, transmitting 
1000 nuclear gradients per second (which would be fast enough even for un-mediated haptic quantum chemistry\cite{Vaucher2016}) requires about 18.3\,Mb/s. This is an order of magnitude less than what is realistically available.

The latency can be a much more problematic parameter. For visual applications, the calculation times must be between 10\,ms and 100\,ms, while for haptic feedback, they have to be as low as 1\,ms, if no mediator potential is activated. Even the small latencies of about 0.1\,ms\,--\,1\,ms achievable on a local area ethernet network introduce a significant delay for haptic applications. Hence, in this situation, it is important that the calculations are either carried out on the same machine on which the results are displayed, or that a special, low-latency network such as InfiniBand is employed. For purely visual feedback, however, even the comparatively high latencies on the internet will usually be unproblematic. Only in cases in which the latency exceeds about 50\,ms, this will lead to a noticeable delay for the user.

\subsection{Key Design Principles and Capabilities of \textsc{Sparrow}}

\textsc{SCINE Sparrow} was originally designed with interactive quantum chemistry as its primary application case. Hence, the considerations of the previous sections heavily affected its design. \textsc{Sparrow} is written in C++, which is a compiled language that typically results in very fast machine-executable code, since contemporary compilers can take advantage of sophisticated optimization procedures. All linear algebra operations are handled by the library Eigen3, which contains highly optimized routines for these operations. Time-intensive parts of the code are parallelized with OpenMP. While this is beneficial for large molecules, the overhead introduced for small molecules means that single-threaded calculations are often faster. Also, note that for all OMx methods and AIQM1, only the calculation of the nucleus--nucleus repulsion and the formation of the Coulomb and exchange matrices from the two-electron integrals are currently parallelized, so there is little (or even no) speedup to be expected for these methods

In addition, \textsc{Sparrow} contains a range of SCF accelerators such as the direct inversion of the iterative subspace (DIIS) method and extended DIIS as well as a combination of both. Moreover, we proposed two schemes to extrapolate the initial density matrix from that of a previously converged calculation\cite{Muhlbach2016}. In interactive calculations, subsequent structures are typically very similar so that such extrapolation schemes work very well, allowing the SCF iterations to converge in fewer steps (see also below).

For all semi-empirical models implemented, the ground-state electronic energy can be calculated, both in a spin-restricted and unrestricted formalism. For the MNDO-type NDDO and DFTB-type methods, analytical nuclear gradients are available; for these methods, Hessian matrices can be calculated with a seminumerical approach by taking derivatives of the analytic gradients. In addition, excited states can be calculated with the TD-DFTB formalism for DFTB2 and DFTB3.
More complex 
calculations such as structure optimizations can be done, \textit{e.g.}, with SCINE ReaDuct\cite{readuct}, exploiting \textsc{Sparrow} as a backend raw data generator. Currently, \textsc{Sparrow} does not feature any implicit solvation model. \textsc{Sparrow} can be employed as standalone command-line binary or as a library to be accessed from other C++ and iPython programs (see also appendix).

To leverage the full power of interactive quantum chemistry, having at one's disposal an efficient way to carry out quantum chemical calculations is, however, not enough. Rather, it is important to combine the human chemical intuition provided by the user with the power of automated calculations, making sure that no important chemical process is undetected. For this, we introduced the concept of molecular propensity\cite{Vaucher2016b}, which denotes the tendency of a system to change its electronic structure such that the underlying potential energy surface needs to be modified (\textit{e.g.}, by addition or removal of a proton or electron, or by adopting an electronically excited state). One can check whether such processes become important by carrying out in the background automated calculations for, \textit{e.g.}, a different spin multiplicity, a different electronic state, or with an additional proton. As soon as the new energy comes close (or is lower) than the energy of the system currently explored by the user, the user can be made aware of this.

\section{Results}
\label{sec:results}

\subsection{Runtimes of Different Models}

In this section, we report runtimes of our new implementation of OM2, OM3, and AIQM1 methods as well as of the already existing methods PM6 and DFTB3 in \textsc{Sparrow} for the calculation of the ground state electronic energy for linear alkanes ranging from methane to hectane (C\textsubscript{100}H\textsubscript{202}). All timings have been obtained on a computer with two Intel Xeon E5-2667 v2 processors (except otherwise stated), repeating every runtime measurement five times. Computer systems usually run several hundred operating-system-related processes already in an ``idle'' state,
which perturb the runtime measurements. To make sure that no particularly large or small perturbation skews the average timing, the largest and smallest timing values were discarded. Afterwards, the arithmetic mean of the remaining three values was taken. In all cases, the standard deviation was found to be at least one order of magnitude smaller than the corresponding average timing, indicating that all timings show a comparatively high precision. For these runtime measurements, \textsc{Sparrow} was compiled with version 10.1.0 of the GCC C++ compiler\cite{gcc}, linking it against the Intel Math Kernel Library 2021.4.0\cite{mkl} (via Eigen 3.4.0\cite{eigen}) and the Boost library 1.73.0\cite{boost}.
The development version of \textsc{Sparrow} used for the present work is available on Zenodo\cite{sparrow_7b41c57}.

\begin{figure}[H]
\begin{center}
\includegraphics[scale=0.5]{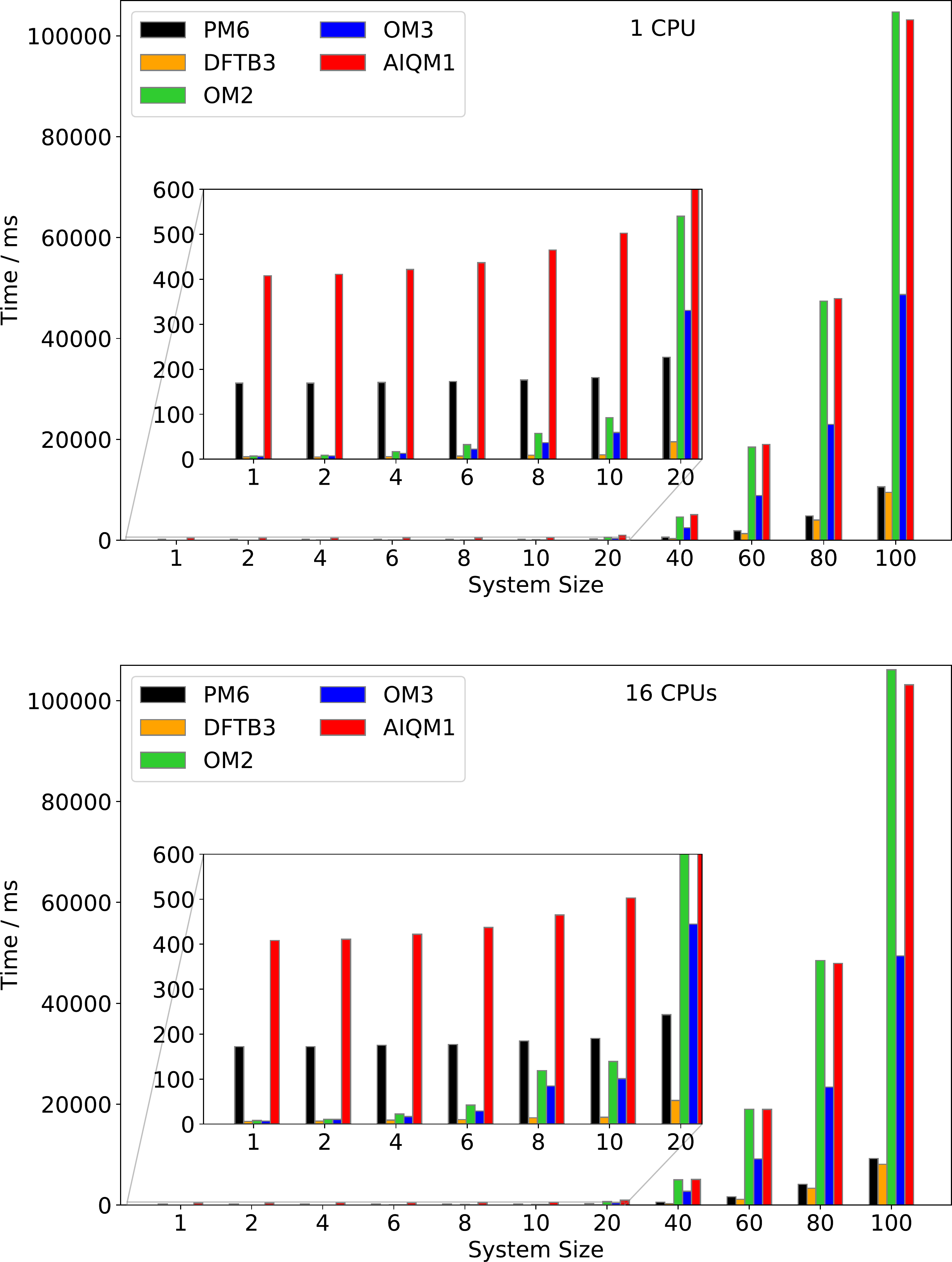}
\end{center}
\caption{\label{fig:timings_1}\small Runtime measurements with PM6, DFTB3, OM2, OM3, and AIQM1 implemented in \textsc{SCINE Sparrow} for linear alkanes of different sizes. The number given on the abscissa specifies the number of carbon atoms in the corresponding alkane. The top part depicts the runtime when utilizing a single processor core, the bottom part shows the runtimes when using 16 CPU cores.}
\end{figure}

In the top part of Fig.~\ref{fig:timings_1}, we show the timings for PM6, DFTB3, OM2, OM3, and AIQM1 as obtained for methane, ethane, butane, hexane, octane, decane, icosane, tetracontane, hexacontane, octacontane, and hectane, measured on a single central processing unit (CPU) core. As one can see, the individual models are associated with different runtimes. For small systems, DFTB3 is consistently the fastest method, even though OM2 and OM3 need only slightly more time to finish the calculation. Starting with hexane, however, for both OM2 and OM3 the time needed to calculate the electronic energy increases rather dramatically, showing a much steeper scaling with system size than DFTB3. For C\textsubscript{20}H\textsubscript{42} and larger systems, the calculation time of OM2 is about ten times larger than that of DFTB3. For these systems, OM3 is about twice as fast as OM2.

AIQM1 shows an identical scaling behavior as OM2 and almost identical timings for the large systems. However, for small systems, namely methane to decane, the computation time for AIQM1 is significantly larger than that of any other method. What is more, for the systems up to hexane, this time is roughly constant, suggesting that for small systems, a constant (\textit{i.e.}, system size independent) effect dominates. This is most likely the initialization routine, combined with the fact that for some parts of the algorithm, external binaries are invoked (the spawning of a new process is rather time demanding). PM6 shows a similar, almost constant runtime for small systems up to decane, before increasing with a similar scaling like DFTB3. For the largest systems, PM6 needs only slightly more time than DFTB3.

It is interesting to investigate the effect of activating additional CPU cores when calculating the electronic energy. The most time-consuming parts of \textsc{Sparrow} are parallelized with OpenMP. However, the benefits of this parallelization are only expected to manifest themselves for very large systems, as for small systems, the overhead imposed by the parallelization adversely dominates the runtime: compare the top of Fig.~\ref{fig:timings_1} to its bottom part, which shows the same timings obtained on 16 CPU cores. 

Up to icosane, all models require more time on 16 cores than on a single one. Starting with tetracontane, DFTB3 as well as PM6 show a slight speedup, whereas the other models still do not benefit from the additional CPU cores. However, as we have noted above, the current implementation of OM2, OM3, and AIQM1 have only very little parallelization, so this is expected. One must keep in mind that most modern CPUs do adjust their operating frequency depending on how many CPU cores are utilized. The  Intel Xeon E5-2667 v2 processor chosen for this work features a maximum peak frequency of 4.00\,GHz. This is the frequency we can expect to be in operation when \textsc{Sparrow} runs on a single CPU core. When all cores are utilized, however, the CPU is expected to throttle the frequency to 3.30\,GHz, implying a performance drop of about 17.5\,\% for a single core. Therefore, the maximum speedup expected when utilizing all 16 cores\,---\,considering only the CPU frequency\,---\,is a factor of 13.2. Obviously, the speedup of a factor of only 1.17 as seen for DFTB3 is clearly below this theoretical threshold, implying that the current parallelization of \textsc{Sparrow} has ample room for improvement.

\subsection{Speedup Attainable on Higher-Performance Hardware}

It is instructive to investigate the effect different CPU models have on the calculation times. All timings reported in the previous section have been obtained on a machine with two Intel Xeon E5-2667 v2 processors. This old CPU type released in the second half of 2013 features a base frequency of 3.30\,GHz and a maximum peak frequency of 4.00\,GHz. For comparison, we repeated the timings on a more recent Intel Xeon W-1390P processor released in 2021, featuring a base frequency of 3.50\,GHz and a maximum peak frequency of 5.30\,GHz. Considering only the peak frequencies, one would expect any \textsc{Sparrow} calculation to run about 1.33 times faster on the more modern CPU.

\begin{figure}[H]
\begin{center}
\includegraphics[scale=0.5]{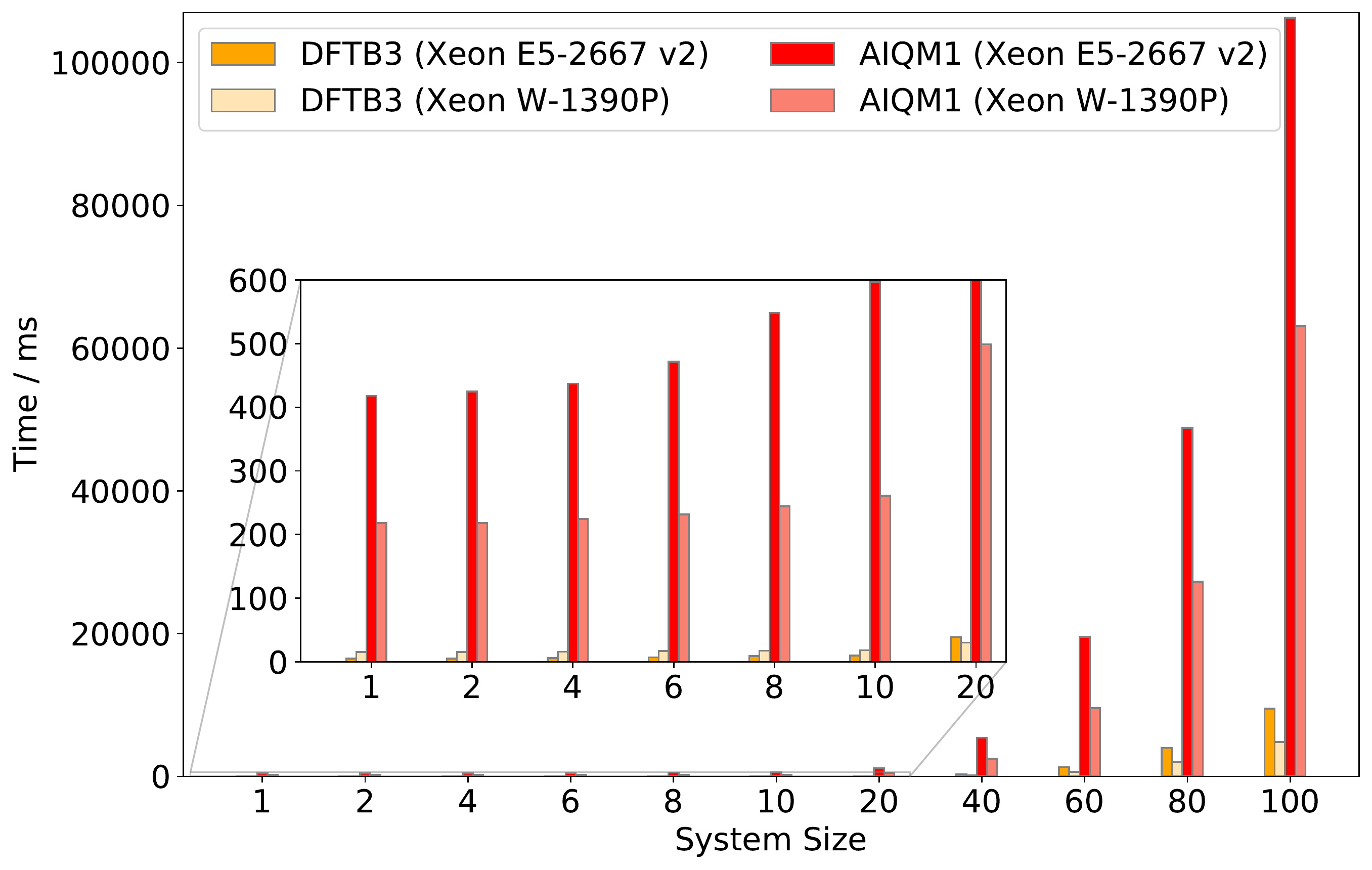}
\end{center}
\caption{\label{fig:timings_2}\small Runtimes of DFTB3 and AIQM1 models implemented in \textsc{SCINE Sparrow} for various linear alkanes on different CPU types.}
\end{figure}

A comparison of the runtimes of DFTB3 and AIQM1 for the series of linear alkanes studied is presented in Fig.~\ref{fig:timings_2}. Here, all calculations were run on a single CPU core. As one can see, for all system sizes calculations with AIQM1 are significantly faster, namely about a factor of 1.8 for the smallest systems and a factor of roughly 1.6 for hectane. This is significantly larger than the theoretical speedup expected based on a comparison of the CPU frequencies alone. This highlights that a multitude of factors affect the calculation runtime besides the processor frequency, such as memory and disk speeds. For our test machine with the more modern CPU, also the rest of the hardware is newer and more performant compared to their counterparts in the older machine. For example, the main memory in the older machine runs at an internal clock frequency of 233\,MHz, achieving a data transfer rate of about 15\,GB/s. In the newer machine, the memory works at 400\,MHz, and the total data transfer rate is almost 26\,GB/s. All these factors combine for a significant total speedup.

While the runtimes of DFTB3 show a similar speedup for the large systems, most curiously, however, this speedup is reversed for the small systems. For methane and ethane, the total time needed to calculate the electronic energy is about a factor of three larger on the more modern computer. Without a more detailed drill-down analysis, it is difficult to unambiguously determine the cause for this reversal. Obviously, the CPU speed as the dominant factor for the computing time is the more important, the larger the systems are. For small systems, the overhead imposed by initialization routines takes a higher percentage of the total runtime than for large systems. During the initialization, \textsc{Sparrow} reads in parameters from disk, among other data. One may speculate that this reading from disk is, for some reason, slower on the newer machine compared to the older one.

\subsection{Comparison to Different Computer Programs}

\begin{figure}[H]
\begin{center}
\includegraphics[scale=0.5]{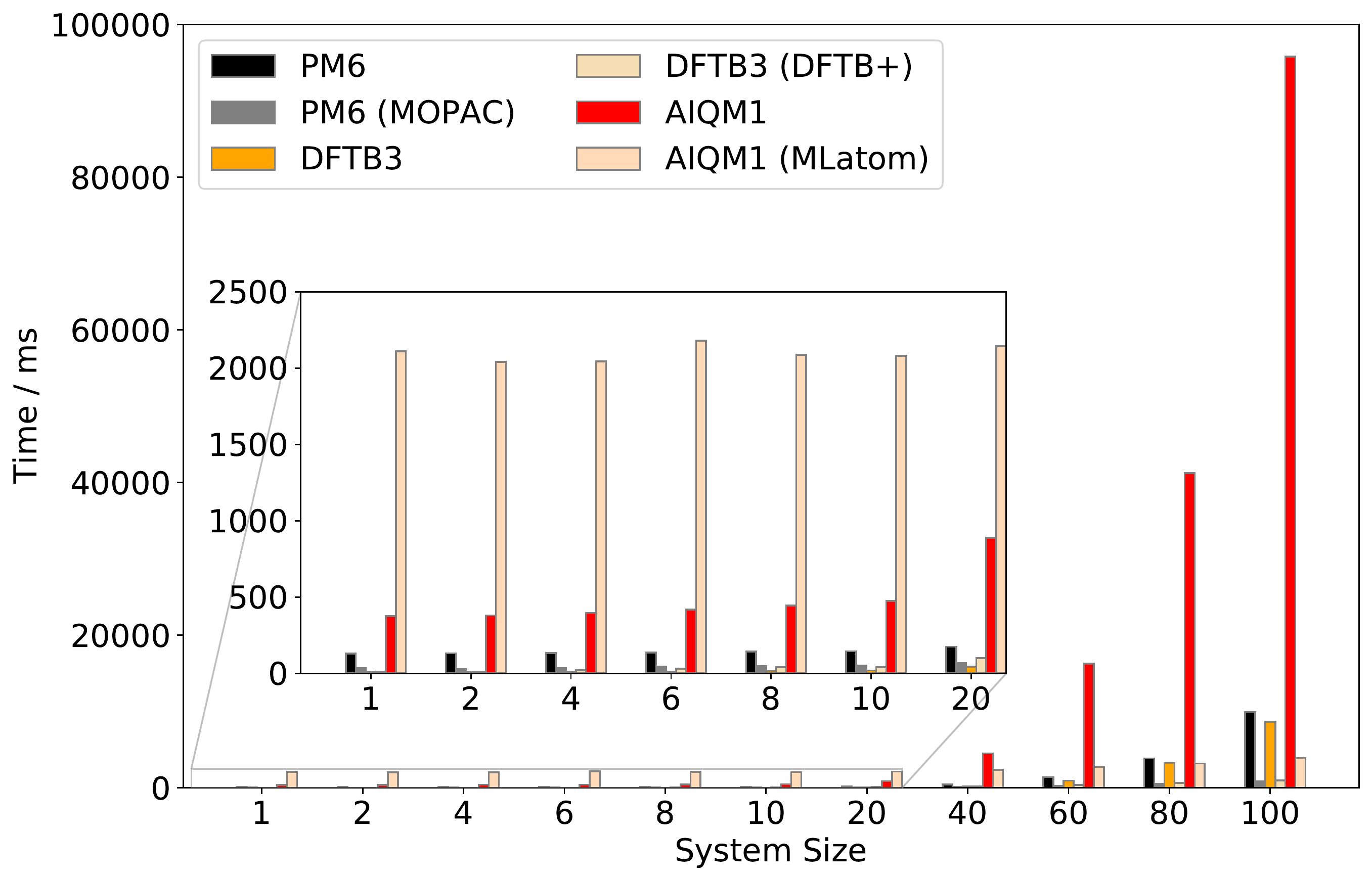}
\end{center}
\caption{\label{fig:timings_3}\small Comparison of runtimes of DFTB3, PM6, and AIQM1 models measured with \textsc{SCINE Sparrow} and compared to the corresponding implementations of \textsc{MOPAC}\cite{mopac}, \textsc{DFTB+}\cite{Hourahine2020}, and \textsc{MLatom} interfaced to the \textsc{MNDO} program, respectively, for various linear alkanes. The timings have been measured on a single core of an Intel Xeon Gold 5222 processor (to obtain these particular timings, \textsc{Sparrow} has been compiled with GCC 7.4.0, the Intel Math Kernel Library 2019.1.144, Eigen 3.3.9, and the Boost library 1.73.0).}
\end{figure}

A comparison of the runtime of \textsc{Sparrow} to that of other programs implementing the same methods is obviously of great interest.
In Fig.~\ref{fig:timings_3}, we present a runtime comparison of PM6, DFTB3 and AIQM1 as implemented in \textsc{Sparrow}, \textsc{MOPAC 2016} (version 19.090L)\cite{mopac}, \textsc{DFTB+ 22.2}\cite{Hourahine2020} and \textsc{MLatom}\cite{MLatomProg,MLatom,MLatom2} (interfaced to the \textsc{MNDO} program\cite{MNDOprog} providing the ODM2* part, \textsc{TorchANI} providing the NN part and \textsc{dftd4} providing the D4 part), respectively. Care has been taken to set numerical thresholds such that convergence settings were as similar as possible for the sake of comparability.

As one can see, the PM6 implementation in MOPAC is consistently faster than the one of Sparrow. Depending on the exact system studied, MOPAC is about a factor of 4 (tetracontane) to 10 (methane, hectane) faster than \textsc{SCINE Sparrow}. For DFTB3, \textsc{Sparrow}'s implementation is somewhat faster for the smaller systems (up to a factor of about 2 for methane) than that provided by DFTB+. However, starting with tetracontane, DFTB+ begins to be faster than \textsc{Sparrow}; for hectane, the implementation of DFTB+ is about a factor of 7 faster. For AIQM1, \textsc{Sparrow}'s implementation also somewhat faster for the smaller systems (up to a factor of about 5 for butane) than that provided by \textsc{MLatom} interfaced to the \textsc{MNDO} program. This is because the file reading and writing process is the most time-consuming part in \textsc{MLatom} for the smaller systems. For larger systems, where calculations of the ODM2* part with the \textsc{MNDO} program dominate, we find that \textsc{MLatom}'s AIQM1 calculations are much faster than by \textsc{Sparrow}, and even faster than PM6 and DFTB3 by \textsc{Sparrow}.

 Despite being slower in some cases, we emphasize that the strong focus on interactive quantum chemistry of  \textsc{Sparrow} has not only motivated the development of ways to reduce the computation time of a individual single point calculation (for which we reported the timings above), but also to accelerate an entire series of such calculations (see also section~\ref{sec:impl_iqc}). Naturally, it can be expected that future developments are likely to also reduce the single-point timings.

\subsection{Implications for High-Throughput Virtual Screening}

We can relate the runtimes discussed above to high-throughput virtual screening as well as interactive quantum chemistry. For this purpose, we first define three different molecule sizes and the typical time required to calculate their ground state electronic energy with \textsc{Sparrow}, \textit{cf.}, Table~\ref{tab:runtimes}.

\begin{table}[H]
\renewcommand{\baselinestretch}{1.0}
\renewcommand{\arraystretch}{1.0}
\caption{\label{tab:runtimes}\small Typical runtimes required by different semi-empirical models of \textsc{SCINE Sparrow} for different molecule sizes. All times are given in ms.}
\begin{center}
\begin{tabular}{l r r r} 
\hline
\hline
Method   & Small (butane-like)  & Medium (icosane-like) & Large (hectane-like)   \\
\hline 
PM6      & 76                   & 100                   & 5300                   \\
DFTB3    & 6                    & 30                    & 4800                   \\
OM2      & 17                   & 290                   & 61800                  \\
OM3      & 13                   & 180                   & 26100                  \\
AIQM1    & 224                  & 500                   & 63100                  \\
\hline
\hline
\end{tabular}
\renewcommand{\baselinestretch}{1.0}
\renewcommand{\arraystretch}{1.0}
\end{center}
\end{table}

For small, butane-like molecules, calculations typically need less than 100\,ms with the exception of AIQM1. For molecules of medium size (comparable to icosane), runtimes vary between 30 and 500\,ms, depending on the method chosen. Large, hectane-like molecules imply runtimes between about 5\,s and one minute. With these runtimes, we can estimate the total amount of molecules one can screen within 10'000 CPU hours, amounting to a small-scale high-throughput virtual screening campaign employing 100 CPU cores for about 4 days (\textit{cf.}, Table~\ref{tab:numbers}).    

\begin{table}[H]
\renewcommand{\baselinestretch}{1.0}
\renewcommand{\arraystretch}{1.0}
\caption{\label{tab:numbers}\small Total number of molecules which can be screened within 10'000 CPU hours by different semi-empirical methods implemented in \textsc{SCINE Sparrow} for different molecule sizes.}
\begin{center}
\begin{tabular}{l r r r} 
\hline
\hline
Method   & Small (butane-like)  & Medium (icosane-like) & Large (hectane-like)   \\
\hline 
PM6      & 473'684'210          & 360'000'000           & 6'792'452              \\
DFTB3    & 6'000'000'000        & 1'200'000'000         & 7'500'000              \\
OM2      & 2'117'647'058        & 124'137'931           & 582'524                \\
OM3      & 2'769'230'769        & 200'000'000           & 1'379'310              \\
AIQM1    & 160'714'285          & 72'000'000            & 570'522                \\
\hline
\hline
\end{tabular}
\renewcommand{\baselinestretch}{1.0}
\renewcommand{\arraystretch}{1.0}
\end{center}
\end{table}

As one can see, very efficient models such as PM6 and DFTB3 can screen several millions of large (hectane-like) molecules in this time frame, while the more involved methods OM2 and AIQM1 can screen more than 500'000 large molecules. For small systems, the throughput rate increases, reaching no fewer than 6 billion small, butane-like molecules with DFTB3 within only 10'000 CPU hours.

\subsection{Implications for Interactive Quantum Chemistry}
\label{sec:impl_iqc}

\begin{figure}[H]
\begin{center}
\includegraphics[scale=0.4]{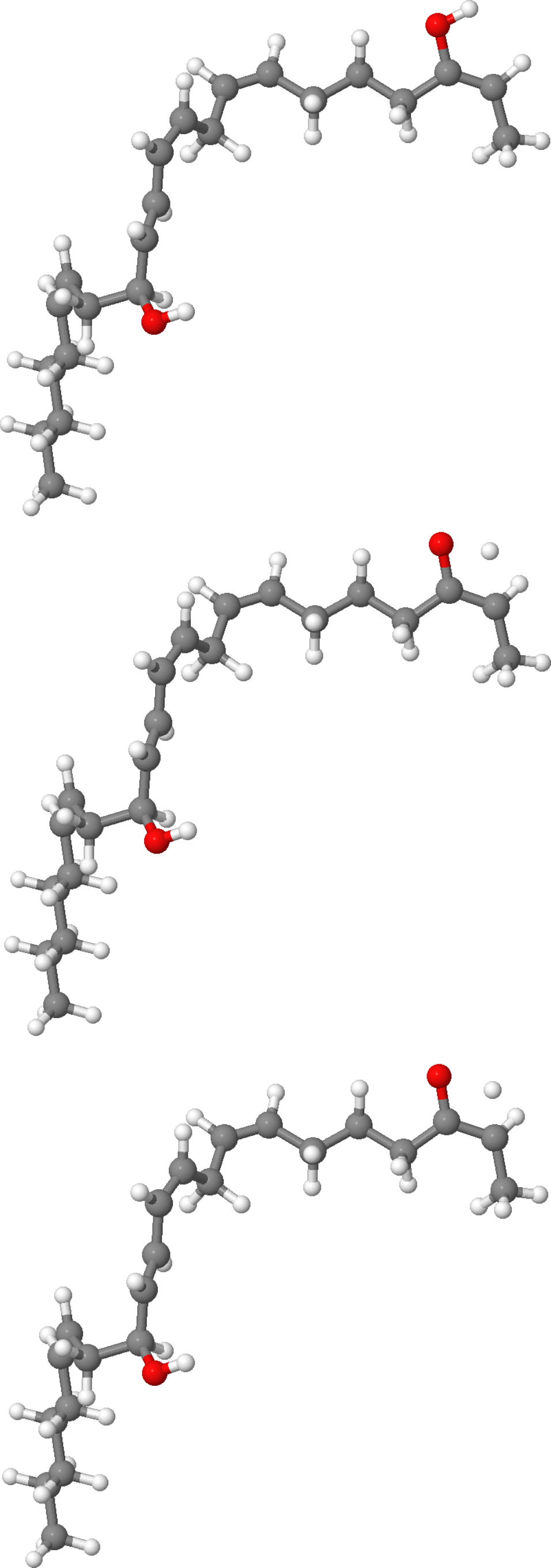}
\end{center}
\caption{\label{fig:trajectory}\small Three snapshots along a trajectory representing a keto--enol tautomerism of a long-chained molecule taken from trajectory T1 of Ref.~\citenum{Bosia2022}. 
Note the similarity of all three structures.}
\end{figure}

Interactive quantum mechanics places a different focus on automated high-throughput calculations: A maximum runtime must not be exceeded, which eventually limits the molecular size that can be studied.
For interactive quantum chemistry, we have seen above that calculation times should not exceed about 34\,ms for a fluent
visual feedback, and for haptic feedback this limit is even lower (depending on additional tricks such as the mediator potential\cite{Vaucher2016}).
Comparing this to the runtimes reported above, we see that the maximum molecule size which can currently be treated interactively is between decane and icosane for the fastest method DFTB3 when only visual feedback is needed (allowing for up to 10\,ms of latency, see above).
However, we have to keep in mind that when a human person is interactively manipulating the atomic coordinates of a molecule, these change in a rather smooth, continuous way. The consequence is that successive structures occurring in an interactive quantum chemistry application are rather similar to each other. Consequently, this can be exploited to accelerate the calculations. In \textsc{Sparrow}, it is possible to use the density matrix of the previous calculation as the starting point for the next calculation\cite{Muhlbach2016}. 
If two structures are rather similar to one another, also their density matrices will be similar, making such a starting guess a good approximation. To showcase this, we demonstrate a series of calculations on a molecular trajectory representing a keto--enol tautomerism of a long-chained molecule. A few snapshots along this trajectory are presented in Fig.~\ref{fig:trajectory} (here, we employ the trajectory called ``T1'' from Ref.~\citenum{Bosia2022}). When calculating all 200 frames of this trajectory with the DFTB3 method implemented in \textsc{Sparrow} independently of each other, the total runtime is 4944.6\,ms. When re-using the density matrix of the previous structure, the runtime drops to 3434.3\,ms, \textit{i.e.}, it is reduced by a factor of 1.4. This is roughly consistent with the decrease of the average number of SCF cycles needed, being 11.7 when the density matrix is not re-used and 7.5 otherwise. A similar speedup by a factor of 1.5 is also observed for the PM6 model. This increase in calculation rate directly translates into larger molecules being amenable to an interactive manipulation. In practice, molecules containing up to thirty heavy atoms are straightforward to deal with in an interactive setting on custom present-day hardware.

We note that for OM3, the speedup observed when injecting the previous density matrix is only a factor of 1.07, even though also in this case, the average number of SCF cycles required for convergence is also reduced by about a factor of 1.5, namely from 14.8 to 10.1. This is evidence for two factors decreasing the computational efficiency of OM3. First, the total number of SCF cycles needed is higher for OM3 compared to DFTB3 or PM6. Second, for OM3, some part of the algorithm outside the SCF routine needs comparatively much time. Clearly, this indicates potential for future optimizations.

For a fluent haptic feedback, the runtimes reported in this work suggest that only very small molecules can be treated. Indeed, the high feedback frequency needed poses a problem also for very efficient semi-empirical methods on contemporary hardware. Therefore, besides a number of algorithmic improvements tailoring the methods in \textsc{Sparrow} to interactive quantum chemistry, also other completely complementary approaches have to be developed. Within the current \textsc{SCINE} framework, the most important of these is the so-called mediator potential, which is a surrogate function modelling essentially a harmonic potential to which the displacement of all atomic nuclei is subjected. The parameters of this potential are obtained from a calculation with \textsc{Sparrow}. Afterwards, this simple function can be evaluated extremely quickly, and restrains the interactive manipulation to the region of the potential energy surface for which a calculation with \textsc{Sparrow} has already been carried out. In the background, semi-empirical calculations are continuously carried out such that the surrogate function can be continuously updated, enabling a fluent manipulation of the system. We refer the interested reader to Ref.~\citenum{Vaucher2016} for more details on this approach.

\section{Conclusions and Outlook}
\label{sec:conclusion}

In this work, we presented the open-source free-of-charge \textsc{SCINE Sparrow} (\url{https://github.com/qcscine/sparrow}), a C++ program implementing a wide range of semi-empirical electronic structure models. We reported a new implementation of the orthogonalization-corrected methods OM2 and OM3 as well as AIQM1, which is an artificial intelligence-enhanced approach based on the OMx model. We discussed key design principles of \textsc{Sparrow}, highlighting how it is especially tailored towards applications in interactive quantum chemistry.

\textsc{SCINE Sparrow} offers a unique collection of semi-empirical approaches broadly ranging from traditional MNDO-type to more advanced orthogonalization-corrected to DFTB-type to state-of-the-art machine learning-improved AIQM1 methods. 
\textsc{Sparrow} is used as backend with MLatom\cite{MLatom,MLatom2} to enable AIQM1 calculations that can be performed on the MLatom@XACS cloud\cite{XACScloud} in a web browser without the need for any software installation, further increasing the accessibility of quantum chemistry.

We then reported the runtimes of a range of semi-empirical models currently implemented in \textsc{Sparrow} for molecules of different sizes, demonstrating the suitability of semi-empirical methods in general as well as their implementation in \textsc{Sparrow} for high-throughput virtual screening, \textit{ab initio} molecular dynamics simulations, as well as interactive quantum chemistry. For screening purposes, semi-empirical methods allow one to screen a huge number of molecules with only modest computational resources. For interactive applications, semi-empirical methods are currently the only way to simulate molecular systems quantum mechanically and on the fly, \textit{i.e.}, without precomputing the potential energy surface.

The runtimes presented in this work also highlight potential directions for the further improvement of \textsc{Sparrow}. For example, the newly implemented methods OM2, OM3, and AIQM1 show a very steep scaling with molecule size. While the time needed by OM2 and OM3 for small molecules is among the shortest of all methods implemented in \textsc{Sparrow}, these methods are about an order of magnitude larger compared to PM6 or DFTB3 for large systems. However, we note that further improvement of the software may increase the efficiency of thee new approaches by an order of magnitude. Furthermore, analytical nuclear gradients are currently not available for OM2, OM3 and AIQM1 in \textsc{Sparrow}. Future work in \textsc{Sparrow} will address these points.

\section*{Author Declaration}
 The authors have no conflict of interest to declare.
 
\section*{Data Availability Statement}
 The data that support the findings of this study are available from the corresponding author upon request.
 
\section*{Acknowledgments}
\label{sec:acknowledgments}

This work was generously supported by the Swiss National Science Foundation (SNSF) through project no.~200021\_182400. P.O.D.~acknowledges funding by the National Natural Science Foundation of China (no.~22003051), the Fundamental Research Funds for the Central Universities (no.~20720210092), and via the Lab project of the State Key Laboratory of Physical Chemistry of Solid Surfaces.

\section*{Appendix: Running \textsc{SCINE Sparrow}}

\textsc{Sparrow} can be used either as a standalone binary via the command line or as a library via C++ or Python. When used as a standalone binary, the command
\begin{verbatim}
sparrow -h
\end{verbatim}
provides a list of all available options. For example, a simple calculation of the electronic energy of a structure in the file \texttt{structure.xyz} with the MNDO method can be achieved with
\begin{verbatim}
sparrow -x structure.xyz -M MNDO
\end{verbatim}
By default, a spin multiplicity of 1 and a charge of 0 are assumed; the calculation is carried out in the spin-restricted formalism. Suppose we want to specify a charge of +1 and a spin multiplicity of 2. For this case, the following options have to be specified:
\begin{verbatim}
sparrow -x structure.xyz -M MNDO -c 1 -s 2
\end{verbatim}
The following example calculates the nuclear gradients as well as the Hessian matrix with the AM1 method:
\begin{verbatim}
sparrow -x structure.xyz -M AM1 -G -H
\end{verbatim}

The following code snippet illustrates the usage of \textsc{Sparrow} via Python; here the electronic energy as well as the nuclear gradients are calculated with the OM3 method:
\begin{verbatim}
import scine_utilities as su
import scine_sparrow

manager = su.core.ModuleManager()
calculator = manager.get('calculator', 'OM3')
calculator.structure = su.io.read('structure.xyz')[0]
calculator.set_required_properties([su.Property.Energy, su.Property.Gradients])

results = calculator.calculate()
print(results.energy)
print(results.gradients)
\end{verbatim}

Further examples and instructions concerning the usage of \textsc{Sparrow} are given in the manual, provided together with the source code of \textsc{Sparrow}\cite{sparrow, sparrow_7b41c57}.

\providecommand{\latin}[1]{#1}
\makeatletter
\providecommand{\doi}
  {\begingroup\let\do\@makeother\dospecials
  \catcode`\{=1 \catcode`\}=2 \doi@aux}
\providecommand{\doi@aux}[1]{\endgroup\texttt{#1}}
\makeatother
\providecommand*\mcitethebibliography{\thebibliography}
\csname @ifundefined\endcsname{endmcitethebibliography}
  {\let\endmcitethebibliography\endthebibliography}{}

\end{document}